\documentclass[a4paper]{aa}
\usepackage{graphicx}
\usepackage{txfonts}
\usepackage{multirow}
\newcommand{\hii}{{\sc Hii}}
\begin{document}
\title{Molecular excitation in the Eagle nebula's fingers}
\author{F. Schuller\inst{1} \and S. Leurini\inst{1} \and C.
Hieret\inst{1} \and K. M. Menten\inst{1}
\and S. D. Philipp\inst{1} \and R. G\"usten\inst{1} \and 
P. Schilke\inst{1} \and L.-{\AA}. Nyman\inst{2}}

\offprints{F. Schuller, schuller@mpifr-bonn.mpg.de}

\institute{Max Planck Institut f\"ur Radioastronomie, Auf
dem H\"ugel 69, D-53121 Bonn, Germany
\and
European Southern Observatory, Alonso de Cordova 3107, Casilla 19001,
Santiago 19, Chile
}

\date{Received xxx / Accepted xxx}

\abstract{The M16 nebula is a relatively nearby \hii ~region,
powered by O stars from the open cluster NGC~6611, which
borders to a Giant Molecular Cloud.
Radiation from these hot stars has sculpted columns of dense
obscuring material on a few arcmin scales. The interface between
these pillars and the hot ionised medium provides a textbook
example of a Photodissociation Region (PDR).}
{To constrain the physical conditions of the atomic and molecular
material with submillimeter spectroscopic observations.}
{We used the APEX submillimeter telescope to map
a $\sim 3'\times3'$ region in the CO $J=$ 3--2, 4--3
and 7--6 rotational lines, and
a subregion in atomic carbon lines. We also observed C$^{18}$O(3--2)
and CO(7--6) with longer integrations on five peaks found in the
CO(3--2) map. The large scale structure of the pillars is
derived from the molecular lines' emission distribution. We
estimate the magnitude of the velocity gradient at the tips
of the pillars and use LVG modelling 
%as well as pointed observations in the C$^{18}$O(3--2) line
to constrain their densities and temperatures.
Excitation temperatures and carbon column densities
are derived from the atomic carbon lines.}
{The atomic carbon lines are optically thin and
excitation temperatures are of order 60~K to 100~K, well consistent
with observations of other \hii ~region-molecular cloud interfaces.
We derive somewhat lower
temperatures from the CO line ratios, of order 40~K. The {\sc Ci}/CO
ratio is around 0.1 at the fingers tips.}{}


\maketitle

\section{\label{sec-int}Introduction}

The interplay of massive stars with their surrounding
interstellar medium, though of critical importance in the
energy budget of a galaxy, is still poorly understood due to
observational limitations, and to the generally large distance
to high mass star forming regions. Located only $1.8 \pm 0.5$
kpc away (Bonatto et al. \cite{ref-bonatto}), the M16
(Eagle) nebula is one of the best templates for detailed
analysis of the environment of high-mass stars.
This nebula is associated with the NGC~6611 star cluster,
which contains more than two dozen stars of spectral type earlier
than B0 (Duch\^ene et al. \cite{ref-duchene}) and hundreds
of lower mass stars and has an age of $1.3\pm0.3$ Myr
(Bonatto et al. \cite{ref-bonatto}).
To the south of this cluster, the radiation from these hot stars
has sculpted columns of dense obscuring material on a few arc\-min
scales, usually referred to as 'fingers' or 'pillars', of which
Hester et al. (\cite{ref-hst}) presented stunning Hubble Space
Telescope (HST) images.
The boundary between the \textsc{Hii} region and the dense
molecular gas provides an archetype of a Photodissociation
Region (PDR).

Indications of present-day star formation near the tips of
the fingers are seen e.g. at infrared (IR) wavelengths
(Thompson et al. 2002).
The fingers have been mapped in the $\rm J=1-0$ line of
CO and isotopologues by Pound (\cite{ref-pound}) with the BIMA
array. White et al. (\cite{ref-white}) used the JCMT 15~m and
OSO 20~m radiotelescopes to map the same region in various
molecular lines and in the submm continuum. They report
CO(2--1) and CO(3--2) peak brightness temperatures above 40~K
and 60~K, respectively, and a factor 2--3 lower temperature
in CO(1--0).

Here we report on new observations in two submillimeter windows,
conducted with the APEX\footnote{This publication is based on data
acquired with the Atacama Pathfinder Experiment (APEX). APEX is a
collaboration between the Max-Planck-Institut f\"ur Radioastronomie,
the European Southern Observatory, and the Onsala Space Observatory.}
12~m telescope. We have mapped an area covering the three pillars in 
CO rotational lines. In addition, we have mapped
the fingertips in the $^3$P$_1$-$^3$P$_0$ and
$^3$P$_2$-$^3$P$_1$ transitions of atomic carbon to probe
the PDR. Details on the observations
are given in Sect.~\ref{sec-obs}. Results of modelling the
dense molecular phase are addressed in Sect.~\ref{sec-molec},
while the properties of the PDR itself are discussed in
Sect.~\ref{sec-pdr}.

\section{\label{sec-obs}Observations}

%\subsection{Observing setup}

We have used the dual
colour FLASH heterodyne receiver (Heyminck et al., this volume) to
simultaneously observe the $\rm J=4-3$ and $\rm J=7-6$ lines of
CO, as well as, also simultaneously, the 492~GHz and 809~GHz
transitions of atomic carbon.
The CO $\rm J=3-2$ line was mapped using the APEX-2a
(Risacher et al., this issue)
facility receiver. With both instruments, the backend
used was the MPIfR Fast Fourier Transform Spectrometer
(Klein et al., this volume), which provides 16~384 spectral
channels over a 1~GHz bandwidth.
Calibration was done by observing the sky,
hot and cold loads and using an atmospheric model.
This provides spectra in antenna temperature unit. 
Efficiencies have been determined on observations of planets
for all receivers (G\"usten et al., this issue) and applied
to the measured signal. The spectra were then converted to the main
beam temperature scale, using beam efficiencies of 0.73
at 345~GHz, 0.60 at 460 and 492~GHz, and 0.47 at 809~GHz.
The overall calibration uncertainty of the
data is estimated to be $\sim$20\%.
The pointing was checked every hour on the nearby sources
Sgr~B2(N) and G10.62--0.38
and was found to be stable within 3$''$. However, parts of
the observations were done at elevations above 80$^\circ$,
where the tracking becomes somewhat less accurate.

We performed position switched observations, with a fixed reference
position to the east of the easternmost finger and at an
offset of $120'', 0''$ relative to our (0,0) position, which is
$\alpha_{2000}=18^{\rm h}18^{\rm m}52\fs 19$,
$\delta_{2000}=-13^{\circ}48'58\farcs 6$. All offsets used
in the following are relative to that position.
We mapped a $2'\times2'$ area with 7$''$ spacing
in both directions in CO(4--3) and CO(7--6), and
$3'\times3'$ with 10$''$ spacing in CO(3--2).
Based on the peaks found in the CO(3--2) map,
five positions were also observed in the (3--2)
transition of C$^{18}$O (see Sect.~\ref{sec-lvg} and
Table~\ref{co-lvg}). Due to moderate weather conditions
for CO(7--6) in the first observing runs, we repeated the
CO(4--3) and (7--6) observations with longer integrations
in April 2006 toward these peak positions.
A limited $40''\times20''$ region covering the tip
of the eastern most finger and a 30$''$ strip on
the tip of the middle finger were mapped in atomic
carbon lines, with 7$''$ spacing in both cases
(Sect.~\ref{sec-pdr}). The log of the observations
is reported in Table~\ref{tab-log-obs}.

\section{\label{sec-molec}Molecular material}

\subsection{Large scale structure}

\begin{figure}[!tp]
\begin{center}
\hspace{9mm}
\resizebox{7.6cm}{!}{\includegraphics{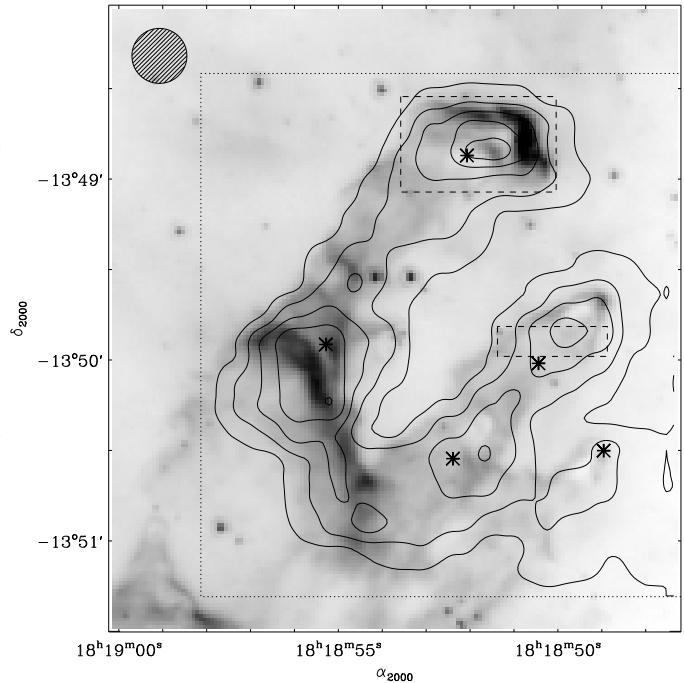}}
\vspace{5mm}
\caption{\label{map_co32}
Contour map of the CO(3--2) emission,
integrated over the +20 to +30~km~s$^{-1}$ V$_{\rm LSR}$ range,
overlaid on a Spitzer 8~$\mu$m image from the GLIMPSE survey
(Benjamin et al.~\cite{ref-glimpse}). Contours correspond to
30, 60,... 180 K~km~s$^{-1}$ in integrated T$_{\rm MB}$.
The APEX beam at 345~GHz is shown in the upper left corner.
The dotted line shows the limits of our CO(3--2) data.
The asterisks show the positions where we ran LVG modelling
(Sect.~\ref{sec-lvg}) and the frames drawn in dashed lines delineate
the areas mapped in atomic carbon lines (Sect.~\ref{sec-pdr}).
}
\end{center}
\end{figure}

\begin{figure*}[!htp]
\begin{center}
\raisebox{1cm}{\resizebox{9cm}{!}{\includegraphics{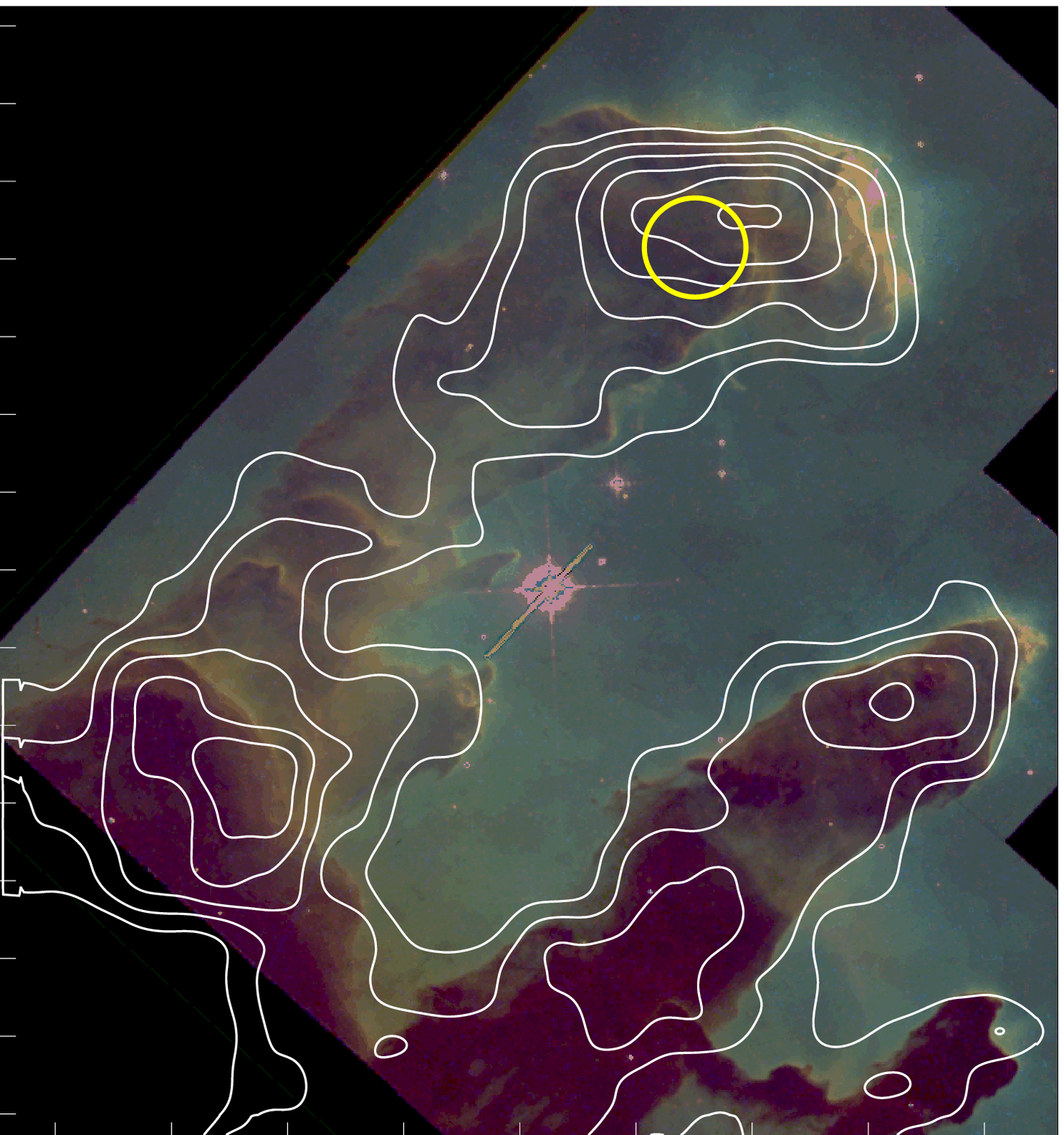}}}
\hspace{5mm}
\resizebox{3.5cm}{!}{\includegraphics{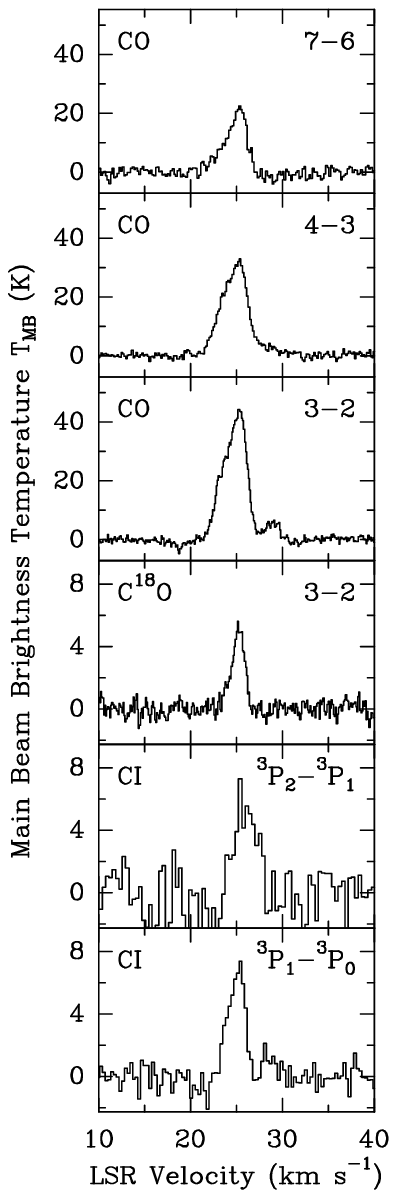}}
%\vspace{5mm}
\caption{\label{map_hst}
\textit{Left hand side}: Contour map of the CO(4--3) emission,
integrated over 20--28~km~s$^{-1}$, overlaid on an HST composite
image ([{\sc O iii}], H$_\alpha$ and [{\sc S ii}] filters)
from Hester et al. (\cite{ref-hst}). Contour levels are shown for
$\int {\rm T_{MB}} dv$ from 40 to 165~K~km~s$^{-1}$ in steps
of 25~K~km~s$^{-1}$.
The circle marks the size of the APEX beam for the CO (4--3) line
and is placed at the position toward which the spectra in the
right hand side of this figure were taken.
\textit{Right hand side, top to bottom}:
Spectra of the CO (7--6), (4--3), (3--2), the C$^{18}$O (3--2)
and the {\sc Ci} $^3$P$_2$-$^3$P$_1$ and
$^3$P$_1$-$^3$P$_0$ lines taken toward the $-10'',0$
offset position marked in the left part of this figure.}
\end{center}
\end{figure*}

A map of the CO(3--2) emission is shown superimposed on 
an 8~$\mu$m Spitzer image in Fig.~\ref{map_co32}. The
8~$\mu$m emission mostly comes from small dust grains
and polycyclic aromatic hydrocarbons (PAHs)
excited by the incoming radiation. It traces
the edge of the dusty structure, while the molecular
emission peaks somewhat deeper in the columns, as
expected from a transition from ionised to atomic
to molecular medium. The integrated intensity in the
CO(4--3) line is shown overlaid on the HST data from Hester
et al. (\cite{ref-hst}) in Fig.~\ref{map_hst}.
Given the pointing uncertainties mentioned above,
the structure of the CO(4--3) agrees remarkably well
with the columns of dust seen in the HST image.

\begin{table}[!htp]
\caption{\label{tab-log-obs}Log of the observations. The T$_{\rm sys}$
column gives typical single side band system temperatures. }
\begin{center}
\begin{tabular}{lll@{ }c@{ }l}
\hline
\hline
Transition & Frequency & T$_{\rm sys}$ & & Date of observation \\
 & [GHz] & [K] & & \\
\hline
CO(3--2)        & 345.796 & 230  & & 17--18 Nov. 2005 \\
C$^{18}$O(3--2) & 329.331 & 330  & & 18 Nov. 2005 \\
CO(4--3)        & 461.041 & 800  & \multirow{2}*[0pt]
{\Huge{$\left. \right\}$}} & 28 Jul. 2005 + \\
                &         &      & & 1 Oct. 2005 + \\
CO(7--6)        & 806.652 & 2500 & & 7-8 Apr. 2006 \\
{\sc Ci} $^3$P$_1$-$^3$P$_0$ & 492.161 & 850 & \multirow{2}*[1pt]
{\LARGE{$\left. \right\}$}} & 02 Aug. 2005 +\\
{\sc Ci} $^3$P$_2$-$^3$P$_1$ & 809.342 & 2000 & & 04--05 Oct. 2005 \\
\hline
\end{tabular}
\end{center}
\end{table}

The 3-dimensional kinematics of the region has been discussed
e.g.~by Pound (\cite{ref-pound}). Our data are fully consistent
with the picture that they derived. In particular, our data
confirm that the CO emission along the easternmost column
arises from two distinct cores. We also observed velocity
gradients (see below) with position angles mostly pointing
to the nearby O stars.

\subsection{Dense cores}

On the tip of the eastern most column, we measure
a difference in V$_{\rm LSR}$ of 2.15$\pm$0.07~km~s$^{-1}$
between offset positions (+10$''$,+20$''$) and (0,-10$''$)
in the CO(3--2) line. This corresponds to a velocity gradient
of 7.8$\pm$0.3~km~s$^{-1}$~pc$^{-1}$, in  good agreement
Pound (\cite{ref-pound}), who found a value of
7.1~km~s$^{-1}$~pc$^{-1}$ when scaled to the distance of 1.8~kpc
used in the present paper.

We derived excitation temperatures by correcting the
main beam temperatures for the Rayleigh-Jeans approximation,
assuming that the CO lines are optically thick and that the
emission is filling the main beam.
%\begin{equation}
%{\rm T_{ex}} = \frac{h \nu}{k \; \ln{(1 + h \nu / k {\rm T_A^*})}}
%\end{equation}
%where we have neglected the radiation from the cosmic background.
At position offset (-10$''$,0), the three
transitions of CO give similar temperatures around
54$\pm$7~K, suggesting that the molecular material
is close to thermal equilibrium.

%\subsection{\label{sec-lvg}LVG modelling}
\subsection{\label{sec-lvg}Gas temperature and column density}

To determine the physical parameters at different positions in the
fingers, we used the CO observations and a spherically symmetric Large
Velocity Gradient (LVG) statistical equilibrium code, with the cosmic
background as only radiation field.  In order to compare the
CO(7--6)  and CO(4--3) observations with the LVG results,
we smoothed the CO(7--6) data to the spatial resolution of the
CO(4--3).
%To reduce the noise level in the spectra,
%we resampled them to a resolution of 0.5~${\rm km~s}^{-1}$.

\begin{table}[h!]
\caption{Results of the CO lines modelling. Cols. 4 and 5 give
the gas kinetic temperatures, as derived by comparing the observed
T$_{\rm MB}$ of CO(4--3) and (7--6), respectively, with the LVG models.
Offsets are relative to the (0,0) position given in Sect.~\ref{sec-obs}.}
\label{co-lvg}
\begin{center}
\begin{tabular}{cllll}
\hline\hline
Offset & $v_{\rm lsr}$ & N(CO) &
T$_{\rm{CO}(4-3)}$ & T$_{\rm{CO}(7-6)}$  \\
%T$^{\rm{CO}(4-3)}_{\rm kin}$ & T$^{\rm{CO}(7-6)}_{\rm kin}$  \\
$\lbrack ''$,$''$] & [km~s$^{-1}$] & [cm$^{-2}$] & [K] & [K] \\
\hline
-10,0    & 23.6 & $\le 1\times 10^{17}$             & 45$\pm 7$ & \\
-10,0    & 25.4 & $8\times 10^{17}-1\times 10^{19}$ & 54$\pm 8$ & 56$\pm 9$\\
+40,-60  & 24.6 & $8\times 10^{17}-6\times 10^{18}$ & 40$\pm 6$ & 42$\pm 6$\\
-30,-70  & 22.4 & $6\times 10^{17}-1\times 10^{19}$ & 50$\pm 9$ & 57$\pm 7$\\
-50,-100 & 21.3 & $6\times 10^{17}-1\times 10^{19}$ & 30$\pm 4$ & 32$\pm 6$\\
0,-100   & 21.4 & $2\times 10^{17}-6\times 10^{18}$ & 51$\pm 9$ & 47$\pm 7$\\
\hline
\end{tabular}
\end{center}
\end{table}

From the CO(3--2)/C$^{18}$O(3--2) line ratios, we estimated the
optical depths of the CO(3--2) lines, and derived CO column densities
by comparing with LVG simulations.
In this calculation, we used $^{16}$O/$^{18}$O=407,
as derived from the relation given by Wilson \& Rood (\cite{ref-wilson})
as a function of galactocentric distance, with d$_{GC}$ = 6.3~kpc
(Bonatto et al. \cite{ref-bonatto}). 
For positon (-10$''$,0$''$), the non-detection of C$^{18}$O(3--2)
at $v_{\rm LSR} = 23.6$~km~s$^{-1}$  gives an upper limit to the CO
column density of 10$^{17}$~cm$^{-2}$. For the other positions, the LVG
modelling gives a range of possible column densities, as reported
in Table~\ref{co-lvg}; our results agree with the average CO column
density derived by Pound (\cite{ref-pound}) and with the C$^{18}$O
column densities of White et al. (\cite{ref-white}).

\begin{table*}[htbp]
\caption{\label{tab-tex-ci}Peak main beam temperatures
of the [{\sc Ci}] lines. The excitation temperatures and line
opacities are computed after Zmuidzinas et al.~(\cite{ref-zmuid}),
using the T$_{\rm MB}$ scale for both lines. Typical uncertainties
are of order -20~K to +30~K for T$_{\rm ex}$, and $\pm$0.1 for
opacities (see text). Column densities are derived from
T$_{\rm ex}$ and $\tau$(1--0) after Keene et al. (\cite{ref-keene}). Offsets
are relative to the (0,0) position (Sect.~\ref{sec-obs}).}
\begin{center}
\begin{tabular}{llllllllll}
\hline
\hline
\multicolumn{2}{c}{Offset} &
T$_{\rm MB}$(1--0) & v$_{\rm LSR}$ &
T$_{\rm MB}$(2--1) & v$_{\rm LSR}$ &
T$_{\rm ex}$ & $\tau$(1--0) & $\tau$(2--1) & N({\sc Ci}) \\
$\lbrack''$] & [$''$] & [K] & [km~s$^{-1}$] &
[K] & [km~s$^{-1}$] & [K] & & & [cm$^{-2}$] \\
\hline
-21 & 0 & 10.8 & 25.37 (0.02) & 11.6 & 25.94 (0.06) &
61  & 0.24 & 0.31 & 3.2$\times10^{17}$ \\
-14 & 0 & 12.2 & 25.54 (0.03) & 14.3 & 25.76 (0.09) &
71  & 0.23 & 0.31 & 3.8$\times10^{17}$ \\
 -7 & 0 &  9.2 & 25.60 (0.06) & 10.4 & 25.70 (0.17) &
66  & 0.18 & 0.24 & 2.9$\times10^{17}$ \\
  0 & 0 &  7.2 & 25.75 (0.06) &  9.7 & 25.52 (0.10) &
93  & 0.09 & 0.14 & 1.8$\times10^{17}$ \\
-35 &-63& 10.3 & 23.10 (0.06) & 14.3 & 22.95 (0.12) &
103 & 0.12 & 0.19 & 2.6$\times10^{17}$ \\
\hline
\end{tabular}
\end{center}
\end{table*}

Using these values for the CO column densities, we further constrained
the gas properties by running LVG calculations for several densities
and temperatures. We found that the CO(7--6) and (4--3) lines are thermalized
for densities higher than a few 10$^4$~cm$^{-3}$. 
Both lines are optically thick at the column densities that we computed
(Table~\ref{co-lvg}); therefore, the observed T$_{\rm mb}$ give direct measures
of the kinetic temperature of the gas. Other authors (Pound \cite{ref-pound},
White et al. \cite{ref-white}) reported densities of
$3-5 \times 10^4$~cm$^{-3}$ in the fingers, with higher values
($\sim2 \times 10^5$~cm$^{-3}$) at the fingertips.
Thus, the temperatures reported in Table~\ref{co-lvg} are derived from
LVG models with densities above $10^4$~cm$^{-3}$. For lower densities,
they would correspond to lower limits to the true kinetic
temperature of the gas. Our estimates are computed by assuming
beam filling factors of 1.
The temperatures we derived do not change significantly over the range of
column densities computed for each position and the variation is, in any case,
within the error bars, corresponding to a calibration uncertainty of 20\%
for each line.

\section{\label{sec-pdr}Atomic carbon lines}

\begin{figure}[!bp]
\begin{center}
\resizebox{8.8cm}{!}{\rotatebox{270}{\includegraphics{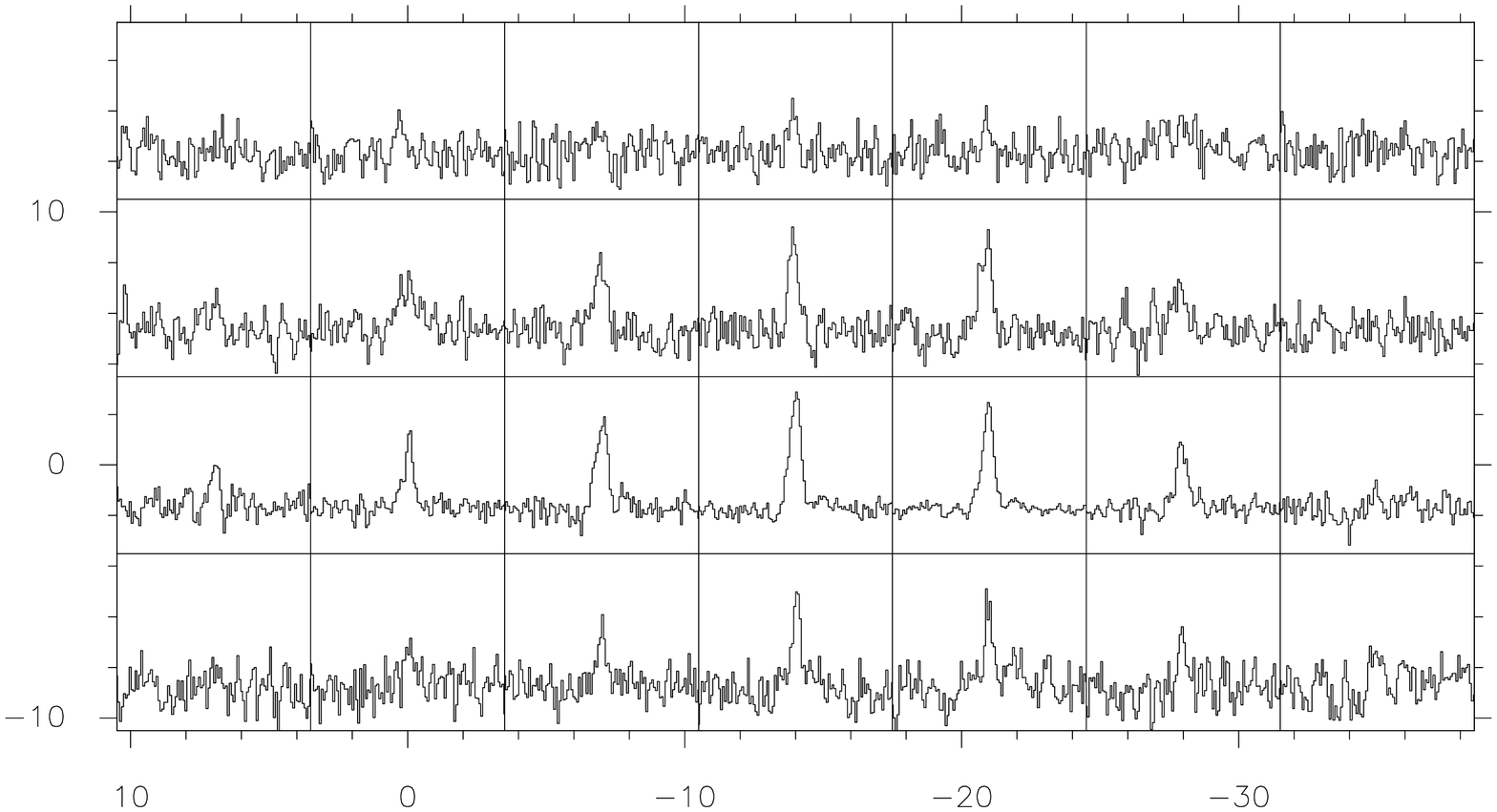}}}
\caption{\label{map_ci}Map of the atomic carbon line at 492~GHz
covering the edge of the northernmost finger. Axes show the offsets
(in arcsec) with respect to the (0,0) position (Sect.~\ref{sec-obs}).
Each spectrum is
shown over the 10--40~km~s$^{-1}$ range in V$_{\rm LSR}$, with a
vertical scale going from -5 to +15~K in T$_{\rm MB}$.
Excitation temperatures derived from the 492~GHz and 809~GHz
transitions are given in Table~\ref{tab-tex-ci}.}
\end{center}
\end{figure}

In the north-south direction, [{\sc Ci}] 492 GHz extends over at
least $\sim$30$''$, or 0.25~pc at the distance of M16 (see map in
Fig.~\ref{map_ci}). Using Eqs.~(7) and (8) in Zmuidzinas et
al.~(\cite{ref-zmuid}), we derived excitation temperatures
and opacities for both [{\sc Ci}] lines, using the measured main beam
temperatures for both transitions. This assumes a uniform excitation
temperature, and that the levels are thermally populated.
Since both lines were observed simultaneously, systematic errors
should cancel out in this analysis, so that using a 10\% uncertainty
for each measured T$_{\rm MB}$ is reasonable. The resulting
uncertainties are of order -20 to +30~K for T$_{\rm ex}$ and
$\pm$0.1 for opacities.

The results are
shown in Table~\ref{tab-tex-ci} for the central part of the
map covering the tip of the first finger, and for offset position
(--35$''$,63$''$) at the tip of the second finger,
where we have good S/N spectra in both transitions.
We find optically thin lines, with opacities in the range
0.1--0.3, and excitation temperatures in the range
60--100~K, i.e. in the upper part of the temperature
range computed by Zmuidzinas et al.~(\cite{ref-zmuid})
in various sources.
We also derived excitation temperatures from LVG modelling,
and found similar results, in the range 80--120~K.

We computed {\sc Ci} column densities using the same formalism
as described in the appendix of Keene et al. (\cite{ref-keene}).
They are reported in the last column of Table~\ref{tab-tex-ci}.
The corresponding {\sc Ci}/CO ratios are around 0.1 at the
tips of both fingers.
However, {\sc Ci} and C$^{18}$O emissions are likely to trace
distinct material, and additional observations are considered
to constrain the variations of {\sc Ci}/CO in the region.

\section{\label{sec-ccl}Summary and conclusion}

We have observed the transitions of atomic carbon for the first
time toward the M16 nebula, and derived excitation temperatures
for the atomic material around 60--100~K, similar to the values
found in other \hii ~region-molecular cloud interfaces.
The kinetic temperatures
derived from CO transitions seem somewhat lower, around 40~K
in most positions. This is consistent with the picture where
the atomic carbon traces the PDR at the edge of the dense
molecular columns.
The {\sc Ci} to CO ratio is $\sim$0.1 at the fingers tips.
%, i.e., in the PDR.
Additional observations and comparison with PDR models are
needed to further constrain the atomic to molecular transition.

\begin{acknowledgements}
We are grateful to F. Boone and A. Belloche for their valuable help
in various aspects of the data processing.
\end{acknowledgements}

\end{document}